\documentclass[twocolumn,showpacs,preprintnumbers,amsmath,amssymb,a4paper,superscriptaddress]{revtex4-1}

\usepackage{graphicx}
\usepackage{dcolumn}
\usepackage{bm}

\newcommand {\nc} {\newcommand}
\nc {\ve} [1] {\mbox{\boldmath $#1$}} \nc {\la} {\mbox{$\langle$}}
\nc {\ra} {\mbox{$\rangle$}} \nc {\beq} {\begin{eqnarray}}
\nc
{\eeqn} [1] {\label{#1} \end{eqnarray}}

\nc {\eol} {\nonumber \\} \nc {\half} {\mbox{$\frac{1}{2}$}}

\begin{document}

\title{Coupled $\ell$-wave confinement-induced resonances in cylindrically symmetric waveguides}
\author{P. Giannakeas}
\email{pgiannak@physnet.uni-hamburg.de}
\affiliation{Zentrum f\"{u}r Optische Quantentechnologien, Universit\"{a}t Hamburg, Luruper Chaussee 149, 22761 Hamburg,
Germany,}

\author{F.K. Diakonos}
\email{fdiakono@phys.uoa.gr}
\affiliation{Department of Physics, University of Athens, GR-15771 Athens, Greece,}

\author{P. Schmelcher}
\email{pschmelc@physnet.uni-hamburg.de}
\affiliation{Zentrum f\"{u}r Optische Quantentechnologien, Universit\"{a}t Hamburg, Luruper Chaussee 149, 22761 Hamburg,
Germany,}

\date{\today}

\begin{abstract}
A semi-analytical approach to atomic waveguide scattering for harmonic confinement is developed taking into account all partial waves.
As a consequence $\ell$-wave confinement-induced resonances are formed being coupled to each other due to the confinement.
The corresponding resonance condition is obtained analytically using the $K$-matrix formalism.
Atomic scattering is described by transition diagrams which depict all relevant processes the atoms undergo during the collision.
Our analytical results are compared to corresponding numerical data and show very good agreement. 
\end{abstract}


\maketitle
\section{Introduction}
Reduced dimensionality in degenerate ultracold atomic gases plays a key role for the experimental realization and theoretical investigation of exotic quantum phases such as the Tonks-Girardeau gas \cite{tg}.
Specifically, atomic collisions in waveguides represent a fundamental ingredient for studying the effects of the reduced dimensionality. 
Indeed, in two-body collisions, the confinement implies significant modifications of the scattering properties of the collisional complex. 
Existing theoretical studies on bosonic collisions show that resonant scattering can be induced by the confinement, yielding the so-called confinement-induced resonance (CIR) effect \cite{olshanii98, bergeman03}. 
A CIR emerges when the length scale of the confinement becomes comparable to the $s$-wave scattering length of the colliding bosons and it is interpreted as a Fano-Feshbach-like resonance. 
An important characteristic property is that the effective one-dimensional two-body interaction can be controlled by adjusting the confinement parameters.
 
In recent years, the significant progress of quantum technologies has lead to the experimental observation and exploration of the CIR physics in quasi-one- and quasi-two-dimensional waveguides for both, bosons \cite{kinoshita04,paredes04,haller09,haller10} and fermions \cite{gunter05,frohlich11}.
Additionally, remarkable theoretical studies focus on CIR physics and its extensions such as $p$-wave CIR in spin-polarized fermions \cite{blume04}, dual-CIR \cite{kim06}, and the resonant molecule formation \cite{melezhik09}. 
Furthermore, the concept of CIR physics has been investigated also in the case of quasi-two-dimensional or anharmonic waveguides \cite{shlyapnikov01,pricoupenko06,idziaszek06,peano05, sala11,peng11} and in the case of multichannel or mixed-dimension scattering  \cite{saeidian08, melezhik11,nishida10}.
Higher-partial wave interactions constitute an interesting extension within the framework of CIR physics as it was shown in \cite{gian11}, since they are expected to provide novel many-body phenomena, such as unconventional superconductivity or superfluidity \cite{rey09, deb09, hofstetter02,deb12}. 
Nevertheless, a thorough theoretical treatment is still needed extending the CIR physics beyond $s$- or $p$-wave interactions.

In this work, we present a theoretical framework for atom-atom collisions in the presence of a harmonic waveguide, where all the higher partial waves are properly taken into account either for bosons or fermions.
Mainly, it consists of a generalization of the $K$-matrix approach for quasi-one-dimensional geometries presented in \cite{blume04}.
Our treatment yields mutually coupled $\ell$-wave confinement-induced resonances as well as an analytical relation for the position of all $\ell$-wave CIRs.
Furthermore, converting the ``physical'' quasi-1D $K$-matrix into a Dyson-like form of equations 
allows us to classify, in terms of specific transition diagrams, all possible processes which the two atoms undergo during the collision.
This detailed analysis permits us to illustrate the connection of the transitions in the manifold of the closed channels of the transverse confinement with the Fano-Feshbach scenario for resonant scattering.   
The validity of our method is verified showing an excellent agreement of the obtained analytical results with numerical calculations of the transmission coefficient in a system of two bosons interacting via a Lennard-Jones potential in the presence of harmonic confinement.

Our work is organized as follows.
In Sec. II we discuss the general aspects of $K$-matrix theory, and present the $K$-matrix approach for quasi-one dimensional systems.
Sec. III contains a comparison of the analytical results with the numerical calculations and a corresponding discussion.
Sec. IV provides a brief summary and conclusions.

\section{$K$-matrix representation for quasi-one dimensional systems}

The $K$-matrix is defined as a Caley transformation of the $S$-matrix given by the following relation:

\begin{equation}
 S=\frac{1+iK}{1-iK}.
\label{1}
\end{equation}
The form of eq.(\ref{1}) is such that the hermiticity of the matrix $K$ ensures the unitarity of $S$.
The advantage of the $K$-matrix with respect to the $S$-matrix is that it possesses a simpler pole structure. 
This allows for an analysis of overlapping resonances in a straightforward manner.

Let us now consider collisions of two atoms in the presence of an external potential, which confines the transversal degrees of freedoms.
The transversal confinement is induced by a harmonic potential, which permits a separation of center of mass and relative degrees of freedom.
Thus the two-body scattering problem can be reduced to a problem of a single particle (relative particle) scattered by a potential (interatomic potential) placed at the origin.
The Hamiltonian of the relative particle, expressed in cylindrical coordinates, reads:
\begin{equation}
H=-\frac{\hbar^2}{2\mu}\nabla^2+\frac{1}{2}\mu\omega_{\perp}^2\rho^2+V_{sh}(\mathbf{r}),
\label{2}
\end{equation}
where $\mu$ is the reduced mass of the two colliding atoms, $\omega_{\perp}$ is the transversal confinement frequency and
$V_{sh}(\mathbf{r})$ is a short-range interatomic potential.
We assume that the interatomic and the transversal confinement potential do not depend on the azimuthal angle $\phi$.
Consequently, the corresponding quantum number $m$ is conserved and throughout this paper we will focus on $m=0$.

The relative Hamiltonian imposes two approximate symmetries in different regions of the configuration space.
Near to the origin, $r<r_0$, the interatomic potential $V_{sh}(\mathbf{r})$ of range $r_0$ dominates and therefore the collision process obeys spherical symmetry. 
Thus in this regime, for $r\approx r_0$, the relative particle experiences a free-space collision off the interatomic potential with the total colliding energy $E=\hbar^2k^2/2\mu$. 
This process can be described by the following radial wavefunction:
\begin{equation}
\Psi_{\alpha}(\mathbf{r})=\sum_{\ell}F_{\ell}(r,\theta)\delta_{\ell \alpha}-G_{\ell}(r,\theta)K^{3D}_{\ell\alpha},
\label{3}
\end{equation}
where $\alpha$ labels the linearly independent solutions, $F_{\ell}(r,\theta)$ ($G_{\ell}(r,\theta)$) is the regular (irregular) solution, 
i.e. spherical Bessel $j_{\ell}(r)$ (spherical von Neumann $n_{\ell}(r)$) functions multiplied by the Legendre polynomials $P_{\ell}(\theta)$, and the summation runs over all the even (odd) $\ell$ for the case of bosons (fermions).
$K^{3D}_{\ell\alpha}$ are the elements of the $K$-matrix $\underline{K}^{3D}$ in three-dimensions encapsulating all the scattering information related to the interatomic potential $V_{sh}(\mathbf{r})$.
Due to the short-range character of the interatomic interactions $\underline{K}^{3D}$ is diagonal and its elements are equal to $\tan \delta_{\ell}$, with $\delta_{\ell}$ being the phase shift of the $\ell$-th partial wave.

Additionally, in the asymptotic limit, $\mathbf{|r|}\rightarrow\infty$, the transversal confinement term in eq.(\ref{2}) dominates implying its cylindrical symmetry.
The total colliding energy $E$ can be written as a sum of the transversal energy being determined by the harmonic confinement and the longitudinal energy, which involves to the $z$-component of the momentum of the relative particle, according to the relation $E=\hbar^2k^2/2\mu=\hbar\omega_{\perp}(2n+1)+\hbar^2q_n^2/2\mu$.
In this region the wavefunctions are axially symmetric and can be written as follows:
\begin{equation}
\Psi_{\beta}(\mathbf{r})=\sum_nf_{n}(z,\rho)\delta_{n\beta}-g_n(z,\rho)K^{1D}_{n\beta},
\label{4}
\end{equation}
where $\beta$ labels the linearly independent solutions, $K^{1D}_{n\beta}$ are the elements of the $K$-matrix $\underline{K}^{1D}$ in quasi-one dimension  and $f_n(z,\rho)$ ($g_n(z,\rho)$) is the regular (irregular) solution in the presence of the trap with no interactions. The specific form of the regular and irregular solutions are as follows:
\begin{equation}
  \bigg( \begin{array}{l l}
    f_n(z,\rho)\\
    g_n(z,\rho)\\
\end{array}
\bigg)
= \Phi_n(\rho)\left \{
    \begin{array}{l l}
      \bigg(\begin{array}{l l}
    \cos q_n|z| \\
    \sin q_n|z|  \\
  \end{array} \bigg)\quad \text{for bosons}\\

  \frac{z}{|z|}\bigg(\begin{array}{l l}
      \sin q_n|z|  \\
      -\cos q_n|z| \\
  \end{array}\bigg)\quad \text{for fermions}\\
\end{array}\right.
\label{5}
\end{equation}
where $\Phi_n(\rho)$ are the eigenfunctions of the two-dimensional harmonic oscillator for $m=0$ and the functions in the curly bracket refer to the longitudinal degree of freedom including the symmetrization (antisymmetrization) for the case of bosons (fermions).

So far it is clear that the scattering process of atomic collisions in the presence of a harmonic waveguide is characterized by two distinct regions in the configuration space which possess different symmetries.
Thus, the key idea is to perform a frame transformation which will permit the matching of eqs.(\ref{3}) and (\ref{4}) yielding an expression of $\underline{K}^{1D}$ in terms of $\underline{K}^{3D}$.
Obviously, a unitary frame transformation in this case doesn't exist since eqs.(\ref{3}) and (\ref{4}) do not fulfill the same Schr\"{o}dinger equation in complete configuration space.
However, the length scales of the interatomic potential $r_0$ and the harmonic oscillator $a_{\perp}=\sqrt{\hbar/\mu\omega_{\perp}}$ differ by orders of magnitudes, $r_0\ll a_{\perp}$ .
This means that between the two distinct regions of spherical and cylindrical symmetry there is a region where $V_{sh}(\mathbf{r})\approx\frac{1}{2}\mu\omega_{\perp}^2\rho^2\approx0$.
Consequently, for this configuration subspace eqs.(\ref{3}) and (\ref{4}) fulfill approximately the same Schr\"{o}dinger equation, and one can perform a local and non-orthogonal frame transformation.
The idea of local frame transformations was introduced by Harmin and Fano \cite{harmin82, fano81} and extended later on by Greene \cite{greene87}.

As it was shown in \cite{greene87} the above-mentioned local frame transformation does not depend on any additional parameter in order to match the two sets of solutions, eqs.(\ref{3}) and (\ref{4}). 
The only condition which has to be fulfilled is $r_0\ll a_{\perp}$ (separation of scales) yielding the following relations, which connect eqs.(\ref{3}) and (\ref{4}):
\begin{eqnarray}
&f_{n}(z,\rho)&=\sum_{\ell} F_{\ell}(r,\theta)U_{\ell n}\quad \text{and} \nonumber \\ 
&g_{n}(z,\rho)&=\sum_{\ell} G_{\ell}(r,\theta)(U^T)^{-1}_{\ell n},
\label{6}
\end{eqnarray}
where $U_{\ell n}$ is the matrix element of the local frame transformation $U$ given by the relation:
\begin{equation}
U_{\ell n}=\frac{\sqrt{2}(-1)^{d_0}}{a_{\perp}}\sqrt{\frac{2\ell+1}{kq_n}}P_{\ell}\bigg(\frac{q_n}{k}\bigg).
\label{7}
\end{equation}
In eq.(\ref{7}) $P_{\ell}(\frac{q_n}{k})$ is the $\ell$-th Legendre polynomial and the term $d_0$ is $\ell/2$ or $(\ell-1)/2$ for the case of bosons or fermions, respectively.
Note that the local transformation $U$ connects the $\ell$ partial waves with the $n$-modes of the transversal confining potential.
Thus $\underline{K}^{1D}$ with the help of eqs. (\ref{4}) and (\ref{6}) will take following form:
\begin{equation}
 \underline{K}^{1D}=\underline{U}^T\underline{K}^{3D}\underline{U}.
\label{8}
\end{equation}
 In eq. (\ref{8}) the matrix $\underline{K}^{1D}$ involves an admixture of all $n$ channels of the transversal confinement with all $\ell$ partial waves of the interatomic potential for all energies.

In our study we focus on the single-mode regime, where the collision takes place in the ground state of the transversal confinement.
Thus, the total colliding energy is $E=\hbar^2k^2/2\mu=\hbar \omega_{\perp} + \hbar^2q^2_0/2\mu$ yielding one energetically open channel ($n=0$) and $n$ energetically closed channels,
where $E<\hbar\omega_{\perp}(2n+1)$ for $n\neq0$.
Note that the total amount of the open and the closed channels is $N$ and it is related to the principal quantum number according to the relation $N=n+1$.
However, the above-mentioned consideration results in an unphysical asymptotic wavefunction $\Psi_{\beta}(z,\rho)$.
Equation (\ref{4}) is given in terms of $f_n(z,\rho)$ and $g_n(z,\rho)$, where the motion in the $z$-direction is described by $\cos q_n|z|$ and $\sin q_n|z|$ (see eq.(\ref{5})).
Now, the quantity $q_n|z|$ becomes imaginary for $n\neq0$ (closed channels), and consequently the $z$-dependent terms in eq. (\ref{5}) contain exponentially diverging subterms for $z\rightarrow\infty$.
In order to restore the correct boundary conditions of the asymptotic wavefunction one has to ``renormalize`` these divergences.
This can be done within the framework of multichannel quantum defect theory (MQDT) by partitioning the wavefunction in open and closed channel subspaces \cite{greene96,burke11}.
In general the wavefunction can be written as follows:
\begin{eqnarray}
 &\Bigg(&\begin{matrix}
	  \underline{\Psi}_{oo} & \underline{\Psi}_{oc}  \\[0.3em]
	  \underline{\Psi}_{co} &  \underline{\Psi}_{cc} \\[0.3em]
      \end{matrix}
\Bigg)
 \Bigg(\begin{matrix}
	  \underline{Y}_{oo}    \\[0.3em]
	  \underline{Y}_{co}  \\[0.3em]
      \end{matrix}
\Bigg)
= \nonumber \\ 
&\Bigg[&\Bigg(\begin{matrix}
	  \underline{f}_{o} & 0           \\[0.3em]
	  0 &  \underline{f}_{c} \\[0.3em]
      \end{matrix}
\Bigg) 
-
 \Bigg(\begin{matrix}
	  \underline{g}_{o} & 0           \\[0.3em]
	  0 &  \underline{g}_{c} \\[0.3em]
      \end{matrix}
\Bigg)
 \Bigg(\begin{matrix}
	  \underline{K}_{oo}^{1D} & \underline{K}_{oc}^{1D}           \\[0.3em]
	  \underline{K}_{co}^{1D} &  \underline{K}_{cc}^{1D} \\[0.3em]
      \end{matrix}
\Bigg)\Bigg]
 \Bigg(\begin{matrix}
	  \underline{Y}_{oo}    \\[0.3em]
	  \underline{Y}_{co}  \\[0.3em]
      \end{matrix}
\Bigg),
\label{9}
\end{eqnarray}
where ``o'' (``c'') indicates the open (closed) subspace referring to $n=0$ ($n\neq0$), respectively and $\underline{\Psi}$, $\underline{Y}$, $\underline{f}$, $\underline{g}$, $\underline{K}$ are partitioned submatrices dictated by the channel decomposition of the considered problem.
In order to eliminate the diverging terms we choose a linear combination of eq. (\ref{9}) where $\underline{Y}_{oo}=\mathbb{I}$ and $\underline{Y}_{co}=-(\underline{K}_{cc}^{1D}-i\mathbb{I})^{-1}\underline{K}^{1D}_{co}$.
Then eq. (\ref{9}) takes the following form:
\begin{eqnarray}
  \underline{\Psi}^{phys}&\equiv& \underline{\Psi}_{oo}\underline{Y}_{oo}-\underline{\Psi}_{oc}(\underline{K}^{1D}_{cc}-i\mathbb{I})^{-1}\underline{K}^{1D}_{co} \nonumber \\
&=&\underline{f}_o-\underline{g}_o[\underline{K}_{oo}^{1D}+i\underline{K}^{1D}_{oc}(\mathbb{I}-i\underline{K}^{1D}_{cc})^{-1}\underline{K}^{1D}_{co}], ~~~~~~~~~~~~
\label{10}
\end{eqnarray}
where $\underline{\Psi}^{phys}$ is the ``physical'' wavefunction which involves only the open channels since the diverging parts of the closed channels are removed.
Moreover, the contribution of the closed channels during the collision process is imprinted in the $K$-matrix $\underline{K}_{oo}^{1D,~phys}\equiv \underline{K}_{oo}^{1D}+i\underline{K}^{1D}_{oc}(\mathbb{I}-i\underline{K}^{1D}_{cc})^{-1}\underline{K}^{1D}_{co}$. 
The roots of $det(\mathbb{I}-i\underline{K}_{cc}^{1D})$ provide the location of the bound states of the closed channels in the parameter space, which energetically lie in the continuum of the open channel.

\begin{figure}[htb!]
\includegraphics[scale=0.32]{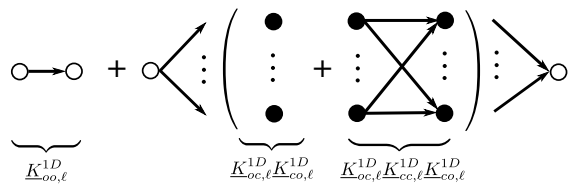}
\caption{A schematic illustration of the transitions into $N$ channels which contribute in the scattering process, corresponding to the $K$-matrix $\underline{K}_{oo, \ell}^{1D,~phys}$. The empty (filled) circles indicate the open (closed) channels, the arrows depict the transitions through the $\ell$-th partial wave, and the dots denote the $N-3$ channels and their corresponding transitions.}
\label{schem}
\end{figure}

As it was shown in \cite{blume04}, for collisions which involve a single partial wave, the matrix elements of the $\underline{K}^{1D} \equiv \underline{K}^{1D}_{\ell}$ are given by the simple relation $\{\underline{K}^{1D}_{\ell}\}_{nn^{\prime}}=(U^T)_{n\ell}\tan \delta_{\ell} U_{\ell n^{\prime}}$, where after the diagonalization and the inversion of $(\mathbb{I}-i\underline{K}_{cc,\ell}^{1D})$ one can obtain $\underline{K}^{1D,~phys}_{oo,\ell}$ in closed form.
However, one should note that $\underline{K}^{1D}_{\ell}$ in this case is a rank one matrix.
Interestingly, one can rewrite $\underline{K}^{1D,~phys}_{oo, \ell}$ in a much simpler form by exploiting this property with the help of \cite{miller81}.
Then $\underline{K}^{1D,~phys}_{oo,\ell}$ reads

\begin{equation}
 \underline{K}_{oo,\ell}^{1D,~phys}=\underline{K}_{oo,\ell}^{1D}+i\underline{K}^{1D}_{oc,\ell}\big( \mathbb{I}+\frac{i}{1-i\gamma}\underline{K}^{1D}_{cc,\ell}\big)\underline{K}^{1D}_{co,\ell},
\label{11}
\end{equation}
where $\gamma=Tr(\underline{K}^{1D}_{cc,\ell})$.
Eq. (\ref{11}) provides us with all the transitions which the two atoms undergo during the collision.
Fig. \ref{schem} illustrates eq. (\ref{11}) as a transition diagram, where the transitions are not direct, but through the $\ell$-th partial wave, which participates as an intermediate process.
Thus, the first term, both in Fig. \ref{schem} and in eq.(\ref{11}) indicates a direct ``open-open'' (oo) channel transition and contains, in particular, the description of potential resonances. 
The second term of eq.(\ref{11}) is related with transitions into closed channels, which can be split into two processes.
As it is shown in Fig. \ref{schem} the second term of eq. (\ref{11}) consists of two transition processes, the first describes an ``open-closed-open'' (oco) channel transition, while the second describes  an ``open-closed'' (oc) channel transition, followed by transitions from {\it{all}} closed channels to {\it{all}} closed channels (cc) and finally a ``closed-open'' (co) channel transition.
Note that a similar idea has been discussed in \cite{alber91}.

The contribution of the transitions into the closed channels during the two-body collision is significantly affected by the factor $1/(1-i\gamma)$, where the parameter $\gamma$ includes all the closed channel transitions through an effective bound state which is supported by all the closed channels and for $\gamma \rightarrow -i$ the bound state coincides with the threshold yielding a resonant collision process.
Indeed, according to MQDT this can be justified since the parameter $\gamma=Tr(\underline{K}_{cc,\ell}^{1D})\rightarrow -i$ is a root of the relation $det(\mathbb{I}-i\underline{K}_{cc,\ell}^{1D})=1-iTr(\underline{K}_{cc,\ell}^{1D})=0$ and as we have mentioned above these roots describe the closed channel bound states at given energy $E$.
Thus, we observe that the $K$-matrix approach provides us with an important insight of CIR physics by linking in a transparent way the physical picture of virtual excitations \cite{olshanii98} and the Fano-Feshbach-like scheme \cite{bergeman03}.
Additionally, an important aspect of the transition diagram (see Fig. \ref{schem}) is that it illustrates in a clear manner the Fano-Feshbach mechanism: the first term represents the continuum (either interacting or non-interacting) and the second term exhibits all the cc transitions, which collectively result to an effective closed channel bound state and its coupling to the continuum.
By applying $\gamma=-i$ for each single $\ell$ partial wave one obtains the condition for resonant scattering of bosons or spin-polarized fermions under strong transverse confinement.
Thus for $\ell=0,1,2$ one will obtain the same resonance condition as in \cite{olshanii98,imambekov10,gian12}, being in agreement with pseudopotential theory.

Up to now we have been focusing on the case of a single partial wave.
By extending the $K$-matrix $\underline{K}^{3D}$ into an $\ell\times\ell$ diagonal matrix, we take into account all the partial waves compatible with the atomic species, i.e. even (odd) $\ell$ for bosononic (fermionic) collisions.
Consequently, by finding the roots of $det\big(\mathbb{I}-i\underline{K}^{1D}_{cc}\big)$ of the extended $K$-matrix, we obtain an expression which provides us with the positions of all the $\ell$-wave confinement-induced resonances coupled to each other.
\begin{equation}
1-i\sum_{\ell=s_0}^{\infty}Tr[\underline{K}_{cc,\ell}^{1D}]-\frac{1}{2}\sum_{\ell=s_0+2}^{\infty}\sum_{\ell^{\prime}=s_0}^{\ell-2}t_{\ell \ell^{\prime}}=0,
\label{12}
\end{equation}
\begin{eqnarray}
&\text{with}&~~t_{\ell \ell^{\prime}}=\tan \delta_{\ell} \tan\delta_{\ell^{\prime}}\sum_{i,j=1}^{\infty}\big(U_{i\ell}U_{j\ell^{\prime}}-U_{j\ell}U_{i\ell^{\prime}}\big)^2\nonumber\\
&\text{and}&~~Tr[\underline{K}_{cc,\ell}^{1D}]=\tan \delta_{\ell}\sum_{n=1}^{\infty}U_{n\ell}^2
\label{13}
\end{eqnarray}

\begin{figure*}[t!]
\begin{center}
\includegraphics[scale=0.30]{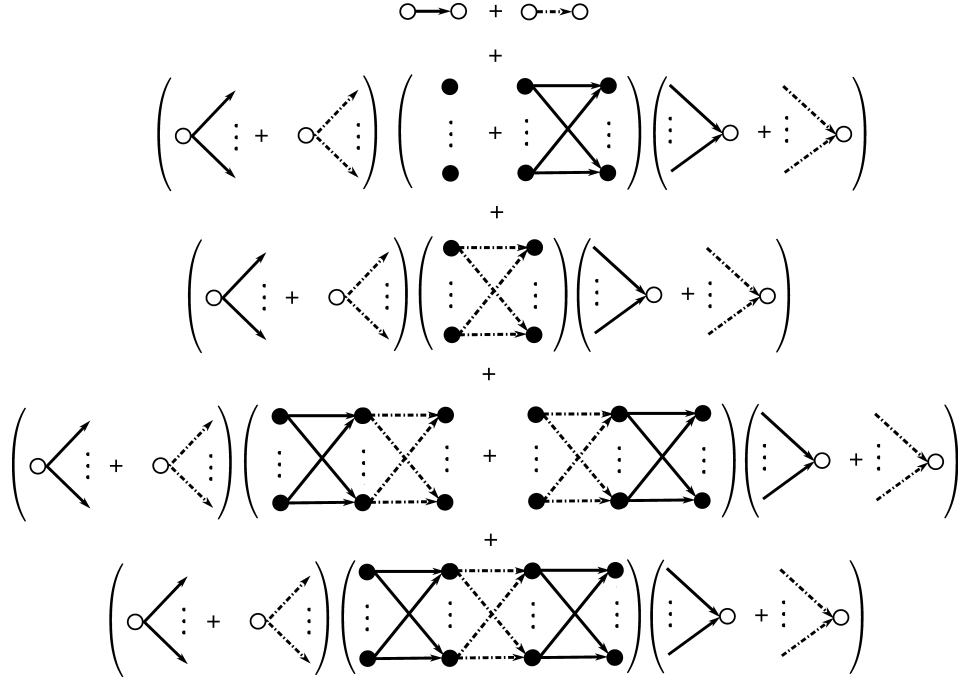}
\end{center}
\caption{A schematic illustration of the transitions into $N$ channels which contribute in the scattering process, for the case of $s$- and $d$-wave. The empty (filled) circles indicate the open (closed) channels, the solid-black (dashed-dotted) arrows depict the transitions through the $s$- ($d$-) partial wave, while the dots denote the $N-3$ channels and their corresponding transitions.}
\label{schem2}
\end{figure*}

Note that for eqs.(\ref{12}) and (\ref{13}) the summation with respect to $\ell$ runs over all the even (odd) $\ell$ with $s_0=0$ ($s_0=1$) for the case of bosons (fermions).
Equations (\ref{12}, \ref{13}) represent the main result of the present investigation.

In order to give some quantitative results in the following we will focus on low energy bosonic collisions ($q_0a_{\perp}\ll1$) considering $s$- and $d$-partial waves. 
Note that a similar analysis, as it is presented below, can be done also for the case of spin-polarized fermions which involve $p$- and $f$- partial waves.

For the scattering of two bosons the matrix $\underline{K}^{1D,~phys}_{oo}$ can be written in terms of $s$- and $d$-wave $K$-matrices in a similar way as in eq. (\ref{11}) yielding the following relation:
\begin{eqnarray}
&\underline{K}&_{oo}^{1D,~phys} = (\underline{K}_{oo,s}^{1D}+ \underline{K}_{oo,d}^{1D})\nonumber\\
&+&i(\underline{K}_{oc,s}^{1D}+\underline{K}_{oc,d}^{1D})(\mathbb{I}+g_1\underline{K}_{cc,s}^{1D})(\underline{K}_{co,s}^{1D}+\underline{K}_{co,d}^{1D})\nonumber\\
&+&ig_2(\underline{K}_{oc,s}^{1D}+\underline{K}_{oc,d}^{1D})\underline{K}_{cc,d}^{1D}(\underline{K}_{co,s}^{1D}+\underline{K}_{co,d}^{1D})\nonumber\\
&+&ig_1g_2(\underline{K}_{oc,s}^{1D}+\underline{K}_{oc,d}^{1D})(\underline{K}_{cc,s}^{1D}\underline{K}_{cc,d}^{1D}\nonumber\\
&+&\underline{K}_{cc,d}^{1D}\underline{K}_{cc,s}^{1D})(\underline{K}_{co,s}^{1D}+\underline{K}_{co,d}^{1D})\nonumber\\
&+&ig_1^2g_2(\underline{K}_{oc,s}^{1D}+\underline{K}_{oc,d}^{1D})\underline{K}_{cc,s}^{1D}\underline{K}_{cc,d}^{1D}\times\nonumber\\
&\times&\underline{K}_{cc,s}^{1D}(\underline{K}_{co,s}^{1D}+\underline{K}_{co,d}^{1D}),
\label{14}
\end{eqnarray}
where $g_1$ and $g_2$ are defined as follows:
\begin{equation}
 g_1=\frac{i}{1-iTr[\underline{K}_{cc,s}^{1D}]},
\label{15}
\end{equation}
and
\begin{eqnarray}
&g_2&=i\big(1-iTr[\underline{K}_{cc,s}^{1D}]\big)/\big(1-iTr[\underline{K}_{cc,s}^{1D}]-iTr[\underline{K}_{cc,d}^{1D}]\nonumber\\
&-&Tr[\underline{K}_{cc,s}^{1D}]Tr[\underline{K}_{cc,d}^{1D}]+Tr[\underline{K}_{cc,d}^{1D}\underline{K}_{cc,s}^{1D}]\big).
\label{16}
\end{eqnarray}

Equation (\ref{14}) is illustrated by the transition diagram in Fig. \ref{schem2} which depicts all the possible transitions that can occur during the collision. 
In this case there are two intermediate transition processes involved, which are related to the $s$- and $d$-partial wave indicated by solid and dash-dotted arrows, respectively.
Additionally, one can observe in Fig. \ref{schem2} that the first and second simple terms describe direct oo transitions via $s$- and $d$-partial waves respectively.
The third (fourth) term describes cc transitions via $s$- ($d$-) partial waves which are directly coupled to the continuum. 
The last two terms of the diagram in Fig. \ref{schem2} describe processes combining cc transitions of both types ($s$- and $d$-partial waves-mediated) coupled to the continuum.
Moreover, in the resonant scattering the factors $g_1$ and $g_2$ control the contribution of the cc transitions via $s$- and $d$-partial waves, respectively. 
One can observe that for $g_1\rightarrow \pm \infty$ and $g_2\rightarrow 0$ the resonant scattering is caused by a s-wave CIR, since the cc transitions through $s$-partial wave become the main resonant mechanism. 
Additionally, for $g_2\rightarrow\pm \infty$ the resonant scattering is caused by a d-wave CIR, since the cc transitions through the $d$-partial wave become dominant, i.e.  are much larger than the contributions of the transitions mediated by the $s$-partial wave.

According to eq. (\ref{12}) the relation which predicts the positions of $s$- and $d$-wave CIR reads
\begin{equation}
 1+c_1(k)\frac{a_s(k)}{a_{\perp}}+c_2(k)\bigg(\frac{a_d(k)}{a_{\perp}}\bigg)^5+c_3(k)\frac{a_s(k)}{a_{\perp}}\bigg(\frac{a_d(k)}{a_{\perp}}\bigg)^5=0,
\label{17}
\end{equation}
where $a_s(k)=-\tan \delta_{\ell=0}(k)/k$, $a^5_d(k)=-\tan \delta_{\ell=2}(k)/k^5$ are the energy dependent $s$- and $d$-wave scattering lengths and $c_1(k)$, $c_2(k)$, $c_3(k)$ are constants related to the Hurwitz zeta functions \cite{abram}, which depend on the total colliding energy according to the following relations:
\begin{eqnarray}
c_1(k) & = & \zeta\bigg[\frac{1}{2},\frac{1}{2}-\bigg(\frac{ka_{\perp}}{2}\bigg)^2\bigg] \nonumber\\
c_2(k)& = & 180 \zeta\bigg[-\frac{3}{2},\frac{1}{2}-\bigg(\frac{ka_{\perp}}{2}\bigg)^2\bigg]\nonumber\\
&+&30(ka_{\perp})^2\zeta\bigg[-\frac{1}{2},\frac{1}{2}-\bigg(\frac{ka_{\perp}}{2}\bigg)^2 \bigg]\nonumber\\
&+&\frac{5}{4}(ka_{\perp})^4\zeta\bigg[\frac{1}{2},\frac{1}{2}-\bigg(\frac{ka_{\perp}}{2}\bigg)^2\bigg]\\
c_3(k) & = & 180 \Bigg\{ \zeta\bigg[-\frac{3}{2},\frac{1}{2}-\bigg(\frac{ka_{\perp}}{2}\bigg)^2\bigg]\times \nonumber\\
&\times&\zeta\bigg[\frac{1}{2},\frac{1}{2}-\bigg(\frac{ka_{\perp}}{2}\bigg)^2\bigg]-\zeta\bigg[-\frac{1}{2},\frac{1}{2}-\bigg(\frac{ka_{\perp}}{2}\bigg)^2\bigg]^2\Bigg\}. \nonumber
\label{18}
\end{eqnarray}
For $q_0a_{\perp}\ll1 \Rightarrow ka_{\perp}\approxeq\sqrt{2}$ the values of these constants are $c_1=-1.46035$, $c_2=-24.3622$ and $c_3=-1.07986$. 

Moreover, the ``physical`` $K$-matrix, $\underline{K}_{oo}^{1D,~phys}$, in the open channels reads
\begin{eqnarray}
&\underline{K}&^{1D,~phys}_{oo}=\frac{1}{q_0a_{\perp}}\frac{1}{1+\frac{a_s(k)}{a_{\perp}}c_1(k)}\bigg[2\frac{a_s(k)}{a_{\perp}}+10\frac{a_d^5(k)}{a_{\perp}^5}\times\nonumber\\
&\times&\frac{[1+(c_1(k)-\frac{c_4(k)}{2})\frac{a_s(k)}{a_{\perp}}]^2}{1+\frac{a_s(k)}{a_{\perp}}c_1(k)+\frac{a_d^5(k)}{a_{\perp}^5}(c_2(k)+c_3(k)\frac{a_s(k)}{a_{\perp}})}\bigg],
\label{19}
\end{eqnarray}
where the constant $c_4(k)$ is given by the relation:
\begin{eqnarray}
c_4(k)&=&12\zeta(-1/2, 1/2-(ka_{\perp}/2)^2)\nonumber\\
&+&(ka_{\perp})^2\zeta(1/2, 1/2-(ka_{\perp}/2)^2).
\label{20}
\end{eqnarray}
In the limit $q_0a_{\perp}\ll1$ the constant $c_4$ takes the value $c_4=-5.41533$.

Equation (\ref{19}) encapsulates all the information of two bosons in a harmonic waveguide involving $s$- and $d$-partial wave scattering. Note that $\underline{K}^{1D,~phys}_{oo}$ possesses two singularities and their positions are given by the following relations:

\begin{equation}
 a_s(k)=-\frac{a_{\perp}}{c_1(k)}~\text{and}~a_d(k)=\sqrt[5]{-\frac{1+c_1(k)\frac{a_s(k)}{a_{\perp}}}{c_2(k)+c_3(k)\frac{a_s(k)}{a_{\perp}}}}~a_{\perp}.
\label{21}
\end{equation}

Singularities occur when the $s$- and $d$-wave scattering length are approximately half of the length of harmonic oscillator $a_{\perp}$ corresponding to $s$-wave and $d$-wave CIR, respectively. 
Furthermore, one observes that the position of the $d$-wave CIR directly depends on the ratio $a_s(k)/a_{\perp}$. 
Interestingly, the second term in the brackets of eq. (\ref{19}), which is related to the $d$-wave CIR, renders the Fano profile of the $s$- and $d$- interfering partial waves, where the nominator describes the width of the $d$-wave CIR depending strongly on the ratio $a_s(k)/a_{\perp}$.

As it was shown in \cite{olshanii98} the scattering of two bosons in the presence of a harmonic waveguide can be mapped onto an effective 1D Hamiltonian of two bosons which interact via an 1D delta potential, $V_{1D}(z)=g_{1D}\delta(z)$, where $g_{1D}=\hbar^2/\mu a_{1D}$, with the effective one-dimensional scattering length being defined as $a_{1D}=\frac{a_{\perp}^2}{2a_s}(-1+1.4603\frac{a_s}{a_{\perp}})$. The coupling strength $g_{1D}$ diverges when $a_{1D}\rightarrow0$ at the position of the $s$-wave CIR.
This idea can be extended in such a way that the $d-$wave CIR is also included in $g_{1D}$ giving $ g_{1D}=\frac{\hbar^2q_0}{\mu} \underline{K}_{oo}^{1D,~phys}$ and 
\begin{eqnarray}
 \frac{1}{a_{1D}}&=& \frac{1}{1+\frac{a_{s}}{a_{\perp}}c_1}\bigg[2\frac{a_s(k)}{a_{\perp}^2}\nonumber\\
&+&10\frac{a_d^5(k)}{a_{\perp}^6}\frac{[1+(c_1-\frac{c_4}{2})\frac{a_s(k)}{a_{\perp}}]^2}{1+\frac{a_s(k)}{a_{\perp}}c_1+\frac{a_d^5(k)}{a_{\perp}^5}(c_2+c_3\frac{a_s(k)}{a_{\perp}})}\bigg]
\label{22}
\end{eqnarray}

\section{Results and discussion}
In this section we compare the analytical results of the previous subsection with the corresponding numerical data.
This comparison is based on Cs atom collisions representing an ideal system for such a study with a rich spectrum of resonances \cite{chin04, mark07}.

First we evaluate the $s$- and $d$-wave energy-dependent
scattering lengths by solving numerically the free space scattering of two Cs atoms interacting via a Lennard-Jones potential,
$V_{sh}(r)=\frac{C_{12}}{r^{12}}-\frac{C_6}{r^6}$. 
The dispersion coefficient $C_6$ has been taken from \cite{chin04}, where the van der Waals length is $\ell_{vdW}=(2\mu C_6/\hbar^2)^{1/4}=202~a_0$ ($a_0$ is the Bohr radius) and $C_{12}$ is a
free parameter, which controls the values of the $s$- and $d$-wave scattering lengths.
Then we use the scattering lengths as an input in eq.(\ref{18}), leading to a parametrization of the $K$-matrix $\underline{K}^{1D,~phys}_{oo}$ in terms of $C_{12}$. 
The corresponding transmission coefficient is calculated by the relation $T=1/[1+(\underline{K}^{1D,~phys}_{oo})^2]$.

Moreover, the numerical simulations of Cs-Cs collisions in the harmonic waveguide are based on \cite{saeidian08, gian11, melezhik12}, where 
 we employ the units $m_{\text{Cs}}/2 = \hbar = \omega_0 = 1$, with $m_{\text{Cs}}$ being 
the mass of the Cs atom and $\omega_0 = 2\pi\times 10$ MHz.
The longitudinal energy is set to $\varepsilon_{\parallel} = 2 \times 10^{-6}$ and the transversal energy is varied within the interval $ 2\times10^{-4} \leq \varepsilon_{\perp} \leq 8 \times 10^{-4}$,
corresponding to the experimentally accessible range $4\pi~\text{kHz} \leq \omega_{\perp} \leq 16\pi~\text{kHz}$ for the waveguide confinement frequency. 
Additionally, the corresponding range of harmonic oscillator length is $5176~a_0 \leq a_{\perp} \leq 2588~a_0$, fulfilling thus the criterion of the harmonic oscillator length being much larger than the range of the interatomic potential, $a_{\perp}\gg \ell_{vdW}$.

\begin{figure}[!htb]
\includegraphics[scale=0.60]{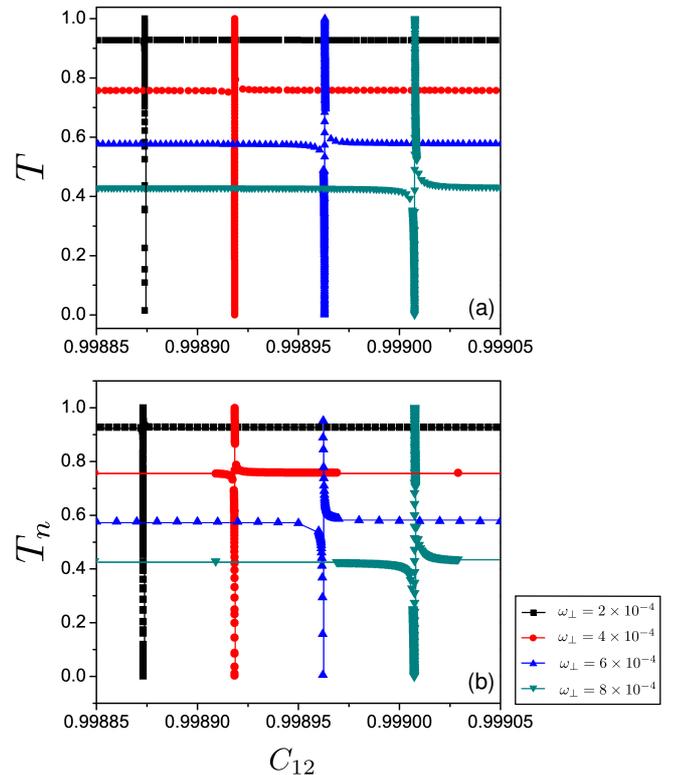}
\caption{(color online) A comparison plot for  (a) analytically calculated transmission coefficient $T$ and (b)
numerically calculated transmission coefficient $T_n$ of $d$-wave CIR for several confinement frequencies $\omega_{\perp}$.}
\label{fig3}
\end{figure}

\begin{figure}[htb!]
\includegraphics[scale=0.65]{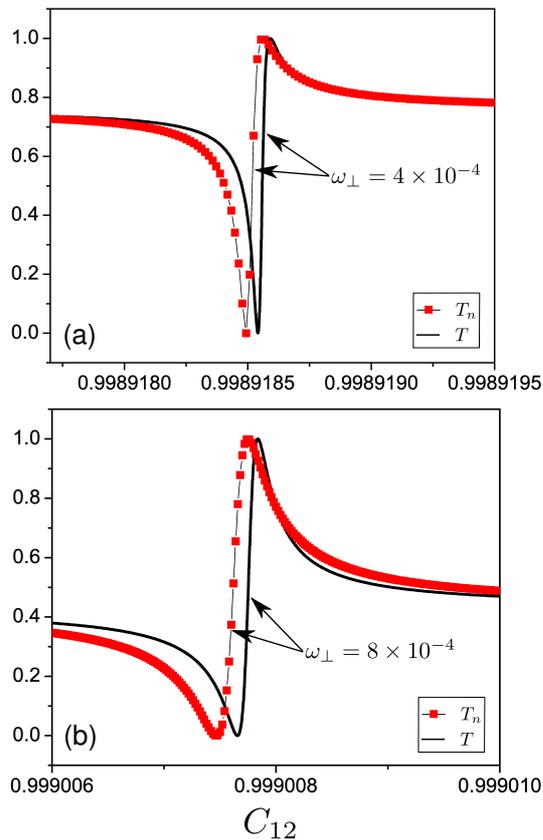}
\caption{ (color online) A high resolution comparison of analytical results (solid line) with numerical data (red squares) for the $d$-wave CIR for two different confinement frequencies, $\omega_{\perp}=4\times10^{-4},~8\times10^{-4}$ in (a),(b) respectively.}
\label{fig4}
\end{figure}

In Fig. \ref{fig3} (a,b) we present the analytically and numerically calculated transmission coefficients, $T$ and $T_n$ respectively, for several confinement frequencies in the vicinity of the $d$-wave CIR. 
Notably, the agreement of the numerical with the analytical calculations is excellent for all confinement frequencies.
Moreover, the analytical results capture the shift of the $d$-wave CIR with increasing confinement frequency and the effect of the strong asymmetric Fano-lineshape of the transmission spectrum $T$.
The latter occurs, due to the interference of $s$- and $d$-wave CIRs, where, as it is indicated in eq. (\ref{19}), the broad $s$-wave CIR serves as a background and the narrow $d$-wave CIR couples to it. 
One should note that this difference in the widths of the two resonances is attributed to the centrifugal term, which is absent in the case of $s$-wave CIR.
Furthermore, we present in Fig. \ref{fig4} a high resolution plot for two confinement frequencies again comparing the analytical with the numerical results.
Even in this fine scale of the $C_{12}$ parameter the analytical calculations follow the numerical ones both qualitatively and quantitatively. 
The small deviations can be explained by the fact that the criterion $a_{\perp}\gg\ell_{vdW}$ is not strongly fulfilled for large confinement frequencies and consequently the analytical results become less accurate.
Indeed, one can observe that these small deviations in Fig. \ref{4} (b), which refer to the case of strong confinement, become less pronounced in the case of intermediate confinement addressed in Fig. \ref{fig4} (a).

\begin{figure*}[t!]
\begin{center}
\includegraphics[scale=0.70]{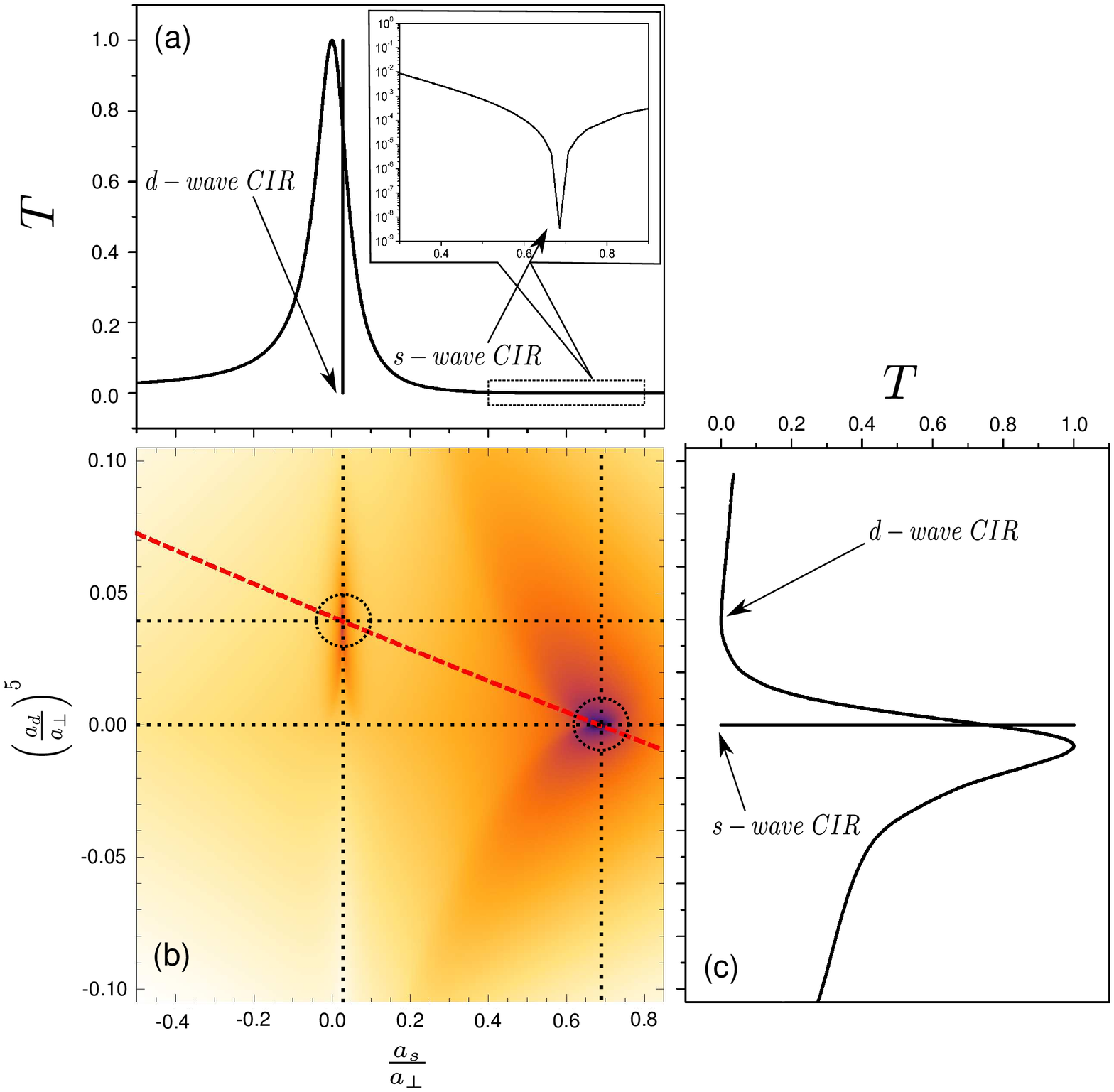}
\end{center}
\caption{(color online) (a) Transmission coefficient $T$ as a function of the $a_s/a_{\perp}$ parameter,
(b) contourplot of the quantity \textbar det$(\mathbb{I}-i \underline{K}^{1D}_{cc})$\textbar~versus the parameters $a_s/a_{\perp}$ and $(a_{d}/a_{\perp})^5$
and (c) transmission coefficient $T$ as a function of the $(a_{d}/a_{\perp})^5$.
In (b) the black dotted lines indicate the position of the resonance with respect to the subfigures (a) and (c), the black dotted circles indicate the positions of the zeros of \textbar det$(\mathbb{I}-i \underline{K}^{1D}_{cc})$\textbar
~and the red dashed line shows the positions of the resonances for an arbitrary short-range interatomic potential. The inset of (a) refers to the dashed boxed area and depicts on a logarithmic scale the second zero of the transmission coefficient ($s$-wave CIR). }
\label{fig5}
\end{figure*}

Fig. \ref{fig5} demonstrates the origin of $s$- and $d$-wave CIR. 
For that purpose we compare the transmission coefficient $T$ (see Figs. \ref{5} (a) and (c)) with the contourplot of the expression \textbar$ \text{det}(\mathbb{I}-i \underline{K}^{1D}_{cc}) $\textbar~ (see Fig. \ref{5}(b)).
The transmission coefficient $T$ and the relation \textbar$ \text{det}(\mathbb{I}-i \underline{K}^{1D}_{cc}) $\textbar~ both depend on the total energy $E=\hbar^2k^2/2\mu$ and the ratios $a_s/a_{\perp}$, $(a_d/a_{\perp})^5$. 
More specific, in Fig. \ref{fig5} (b) for a given energy equal to the total colliding energy $E$ the quantity \textbar$ \text{det}(\mathbb{I}-i \underline{K}^{1D}_{cc}) $\textbar~ has two zeros whose position is indicated in the contourplot by the dark shaded areas (dashed circles), where the positions of the zeros correspond to two different pairs of ratios ($a_s/a_{\perp}$, $(a_d/a_{\perp})^5$).  
The latter means that the closed channels possess two distinct bound states with energy equal to the total energy $E$ at the corresponding values of the pairs ($a_s/a_{\perp}$, $(a_d/a_{\perp})^5$), where one bound state is related to the $s$-wave ($(a_d/a_{\perp})^5\approx0$) and the other one is related to the $d$-wave. 
On the other hand, in Figs. \ref{fig5}(a) and (c)  the transmission coefficient $T$ of the open channel at the same energy $E$ possesses two zeros which correspond to resonant scattering. 
The position of the $d$-wave CIR is in Fig. \ref{fig5}(a) at the point $a_s/a_{\perp}=0.02$ and in Fig. \ref{fig5}(c) at the point $(a_d/a_{\perp})^5=0.04$.
Respectively, the position of the $s$-wave CIR is in Fig. \ref{fig5}(a) (see also the inset plot on a logarithmic scale) at the point $a_s/a_{\perp}=0.68$ and in Fig. \ref{fig5}(c) at the point $(a_d/a_{\perp})^5\approx0$.
Thus, we observe an exact matching of the pairs ($a_s/a_{\perp}$, $(a_d/a_{\perp})^5$) at $T=0$, denoted by the dotted vertical and horizontal lines in Fig. \ref{fig5}(b), with the pairs of ($a_s/a_{\perp}$, $(a_d/a_{\perp})^5$) obtained from the zeros of the relation \textbar$ \text{det}(\mathbb{I}-i \underline{K}^{1D}_{cc}) $\textbar.
The latter means that the $s$- and $d$-wave confinement-induced resonances occur due to the corresponding $s$- and $d$-wave bound states of the closed channels, which couple to the continuum of the open channel. 
Furthermore, due to the harmonic confinement, $s$- and $d$-wave CIR are coupled together rendering interference effects, i.e. a strong asymmetric Fano-lineshape of the transmission coefficient $T$, as it is depicted in Fig. \ref{fig5}(a) and (c). 
Finally, in Fig. \ref{fig5}(b) the red dashed line indicates the position of the $d$-wave CIR for any short-range interatomic potential according to eq. (\ref{17}).

\section{Brief Summary and Conclusions}
We have explored two-body scattering within a harmonic waveguide taking into account all possible partial wave excitations. 
Our analysis is based on the assumption that the range of the interatomic interactions has to be smaller than the length scale of the harmonic confinement, imposing thus two regimes of distinct symmetries, i.e. spherical for $\mathbf{r} \rightarrow 0$ and cylindrical for $\mathbf{r} \rightarrow \infty$.
The expansion over $\ell$-partial waves leads to an extension of CIR physics, where more $\ell\neq0$-wave CIRs emerge being coupled to the broad $s$-wave CIR predicted in \cite{olshanii98}.
Furthermore, we obtain analytically the positions of all $\ell$-wave confinement-induced resonances (eqs.(\ref{12}, \ref{13})). 
This general result refers both to fermionic and bosonic collisions. 
Note that it can be easily extended for more than one open channel taking into account inelastic processes as well.
Additionally, we have demonstrated that the $K$-matrix approach can provide us with a unique insight of the underling physics which scattering in confined geometries can undergo (eqs.(\ref{11}, \ref{14})). 
More specifically, we show that the ''physical'' 1D $K$-matrix, $K_{oo}^{1D,~phys}$, can be expanded in a Dyson-like form of equations, where each term describes all possible transitions which occur during the collision process.
The latter permits us to classify and thoroughly analyze all couplings which are induced by the presence of higher $\ell$-partial waves.
Moreover, this scheme allows  to unravel the connection of the two interpretations for the CIR effect introduced in \cite{olshanii98,bergeman03}, where all the possible transitions through each $\ell$-partial wave into the closed channels  act in a collective way yielding an effective $\ell$-wave bound state supported by all closed channels being coupled to the continuum of the open channel. 
The latter yields an asymmetric Fano-lineshape in the transmission coefficient $T$.
Additionally, we implement the extended analytical model in the case of two-body bosonic collisions in the presence of a harmonic waveguide with the scattering process involving only $s$- and $d$-partial waves, where we observe an excellent agreement between the analytical and numerical results.
Furthermore, we analyze the $d$-wave CIR being coupled to the $s$-wave CIR, where both resonances depict an asymmetric Fano-lineshape in the transmission spectrum.
A similar behavior is expected also for the case of spin-polarized fermionic collisions involving $p$- and $f$-partial waves and their couplings.
Nevertheless,  we should mention that for higher partial waves ($\ell\geq4$) the widths of the corresponding $\ell$-wave CIRs become very narrow, due to the pronounced centrifugal barrier which the atoms have to tunnel through in order to perform a $\ell$-partial wave collision.
Finally, we would like to mention that within the framework of the $K$-matrix approach the origin of the narrow $\ell$-wave resonances in the presence of confinement is a Fano-Feshbach-like.
The latter is not in conflict with the interpretation given in \cite{gian11}, due to the fact that when an $\ell\neq0$ Fano-Feshbach resonance is slightly above the threshold, it can be seen also as a shape resonance \cite{gao09, gao11}. 
This situation occurs also in the considered system because the energy of the maximum of the centrifugal barrier is three orders of magnitude larger than the energy of the ground state of the harmonic confinement, thus the atoms in the $d$-wave Fano-Feshbach-like state tunnel through the barrier exactly as if they would be in a single-channel shape resonant state. 
This complex behavior can be described in a transparent way within the $K$-matrix approach.

\begin{acknowledgements}
The authors are thankful to C.H. Greene and V.S. Melezhik for fruitful discussions.
P.S. acknowledges the financial support by the Deutsche Forschungsgemeinschaft.
\end{acknowledgements}

\end{document}